\pdfoutput=1 

\documentclass[conference,10pt,a4paper]{IEEEtran}
\IEEEoverridecommandlockouts
\usepackage{amsmath}
\usepackage{amsthm}
\usepackage{amsopn}

\usepackage{float}

\usepackage{amssymb}

\usepackage{amsfonts}

\usepackage{amsthm}

\usepackage{graphicx}

\usepackage{booktabs}

\usepackage{caption3} 
\DeclareCaptionOption{parskip}[]{} 
\usepackage[small]{caption}

\usepackage{setspace}
\usepackage{setspace,graphicx,multirow}
\usepackage{subcaption}

\usepackage[ruled,linesnumbered]{algorithm2e}
\usepackage[]{algpseudocode}
\makeatletter
\def\BState{\State\hskip-\ALG@thistlm}
\makeatother
\usepackage{algcompatible}

\usepackage{float}
\usepackage{authblk}
\usepackage{geometry}
\usepackage{cite}

\usepackage{color}
\ifodd 1
\newcommand{\com}[1]{\textbf{\color{blue} (COMMENT: #1)}} 
\else

\newcommand{\com}[1]{}
\fi

\usepackage{enumitem}
\setenumerate[1]{itemsep=0pt,partopsep=0pt,parsep=\parskip,topsep=0pt}
\setitemize[1]{itemsep=0pt,partopsep=0pt,parsep=\parskip,topsep=0pt}
\setdescription{itemsep=0pt,partopsep=0pt,parsep=\parskip,topsep=0pt}

\begin{document}

\bibliographystyle{IEEEtran}
\bstctlcite{IEEEexample:BSTcontrol}

\title{Fast Online Movement Optimization of Aerial Base Stations Based on Global Connectivity Map}

\author{Yiling~Wang,~\textit{Student Member,~IEEE}, Jiangbin~Lyu,~\IEEEmembership{Member,~IEEE}, and~Liqun~Fu,~\IEEEmembership{Senior Member,~IEEE}%
\thanks{The authors are with the School of Informatics, and also the Institute of Artificial Intelligence, Xiamen University (XMU), China, the Shenzhen Research Institute of XMU, China, and the Sichuan Institute of XMU, China. \textit{Corresponding author: Jiangbin Lyu}, (ljb@xmu.edu.cn).}}

\maketitle
\begin{abstract}
Aerial base stations (ABSs) mounted on unmanned aerial vehicles (UAVs) are capable of extending wireless connectivity to ground users (GUs) across a variety of scenarios.
However, it is an NP-hard problem with exponential complexity in  $M$ and $N$, in order to maximize the coverage rate (CR) of $M$ GUs by jointly placing $N$ ABSs with limited coverage range.
The complexity of the problem escalates in environments where the signal propagation is obstructed by localized obstacles such as buildings, and is further compounded by the dynamic GU positions.
In response to these challenges, this paper focuses on the optimization of a multi-ABS movement problem, aiming to improve the mean CR for mobile GUs within a site-specific environment.
Our proposals include
1) introducing the concept of global connectivity map (GCM) which contains the connectivity information between given pairs of ABS/GU locations;
2) partitioning the ABS movement problem into ABS placement sub-problems and formulate each sub-problem into a binary integer linear programming (BILP) problem based on GCM; 
3) and proposing a fast online algorithm to execute (one-pass) projected stochastic subgradient descent within the dual space to rapidly solve the BILP problem with near-optimal performance.
Numerical results demonstrate that our proposed method achieves a high CR performance close to the upper bound obtained by the open-source solver (SCIP), yet with significantly reduced running time.
Moreover, our method also outperforms common benchmarks in the literature such as the K-means initiated evolutionary algorithm or the ones based on deep reinforcement learning (DRL), in terms of CR performance and/or time efficiency.

\end{abstract}

\section{Introduction}\label{sec:introduction}
Unmanned aerial vehicles (UAVs) play an indispensable role in the recent emergence of low-altitude economy, due to their enhanced mobility and decreasing costs.
In particular, one promising application is the employment of UAVs as aerial base stations (ABSs), which could provide timely on-demand wireless connectivity to mobile ground users (GUs) in diverse scenarios, especially when the fixed communication infrastructure is compromised/overloaded\cite{LyuTWCHotspot}. 
A pivotal challenge here lies in determining a suitable placement of $N$ ABSs with restricted coverage range to attain the maximum coverage for $M$ GUs, which is an \textit{NP-hard problem} with exponential complexity in $M$ and $N$ \cite{2017-Lyu-Placement}.
A myriad of efficient heuristic algorithms have been proposed to address this problem, such as the spiral algorithm \cite{2017-Lyu-Placement}, K-means algorithm \cite{2016-Galkin-Deployment}, circle packing theory \cite{2016-Mozaffari-Efficient}, and user-majority based adaptive UAV deployment \cite{8760267}, while it is time-consuming to obtain the optimal solution under large $M$ and $N$.
Moreover, under the typically assumed dominant line-of-sight (LoS) channel or probabilistic LoS/non-LoS (NLoS) channel model\cite{2014-Hourani-Optimal}, the coverage region per ABS is essentially simplified as a uniform disk, which, however, could become irregular when considering site-specific blockages with local LoS/NLoS conditions\cite{2019-Lyu-RadioMap}.
Such \textit{site-specific channel} effects significantly add to problem complexity. 
Finally, the practical scenarios with \textit{GU mobility} further compound the problem by requiring timely adaptation of ABS positions\cite{10373821}.

Regarding the site-specific multi-ABS placement/movement problem, we identify two main streams of research, including model-free and model-based methods.
The model-free methods are represented by the deep reinforcement learning (DRL)-based approach that learns online/offline from interactions with the environment and/or among the ABSs/GUs\cite{2020-Liu-Distributed, 2019-Liu-Reinforcement,2021-Zhang-Three-Dimension, 2020-Qiu-Placement, 2020-Lyu-Codesign}, whereby most works consider generic statistical channel models and only few consider site-specific environments\cite{2020-Qiu-Placement}\cite{2020-Lyu-Codesign}. Even so, the considered network size is typically small/moderate due to the inherent complexity discussed above.
Specifically, DRL-based methods require careful state-action-reward design, and become more difficult to converge due to the curse of dimensionality (with $M$ and $N$, and the number of interaction steps) and non-stationarity of environment (e.g., network dynamics due to GU mobility and on-off activities).

In contrast, the model-based paradigm is represented by radio map (RM)\cite{2019-Lyu-RadioMap} or channel knowledge map (CKM)\cite{CKMtutorial}-based methods to construct/utilize spatial channel distribution, in the context of cellular-connected UAVs\cite{2019-Zhang-RadioMap-Path}\cite{2021-Zeng-Navigation-RadioMap}, or ABSs\cite{10486853}.
The authors in \cite{10486853} assume given RMs obtained via ray-tracing simulations or tomographic measurements, based on which they could relax the NP-hard ABS placement problem into convex ones. Nevertheless, the ABS movement problem to cope with GU mobility is not explicitly studied, where stringent time limit should be imposed on the algorithm.
Finally, regarding model-based methods, other than RM, our recent work in \cite{10373821} propose a deep learning (DL)-based environment emulator that could predict the covered GUs given any number of ABSs/GUs in a site-specific environment. Nevertheless, though fast and accurate during inference, a fair amount of training data and training time is required before the DL model can be on course.
Therefore, in cases where RMs are readily available, e.g., from effectively accumulated historical measurements\cite{2019-Lyu-RadioMap}, from dedicated UAV radio mapping\cite{2023-liu-UAV-Aided}, from fast inference of geometry information\cite{levie2021radiounet}, or from continuous update based on environmental changes\cite{zhen2022radio},
we make another attempt based on RM to achieve fast online ABS movement.

To this end, we first introduce the concept of global connectivity map (GCM), which is an abstracted form of RM that focuses on the connectivity information between any given pairs of ABS/GU locations in a site-specific environment. 
Second, the ABS movement problem is divided into ABS placement sub-problems,
each aiming to maximize the coverage rate (CR) of all GUs in a short time period subject to ABS movement constraints. 
Third, each ABS placement subproblem is formulated as a binary integer linear programing (BILP) problem based on GCM, for which a novel fast online algorithm \cite{li2020simple} is introduced with tailored modifications to fit the problem. 
In particular, We narrow down the search range by considering the ABS mobility model, which helps reduce the computational complexity of the algorithm.
Optimality bounds and complexity analysis are also provided.
Finally, numerical results demonstrate that our proposed algorithm achieves a high CR performance close to the upper bound obtained by the open source solver (SCIP)\cite{BestuzhevaEtal2021ZR}, yet with significantly reduced running time.
Moreover, our algorithm also outperforms common benchmarks in the literature such as the K-means initiated evolutionary algorithm or the ones based on DRL, in terms of CR performance and/or time efficiency.


\section{System Model and Problem Formulation}\label{SectionSystem}
\begin{figure}[!t]
\centering
\includegraphics[width=0.38\textwidth,  trim=5 5 5 5,clip]{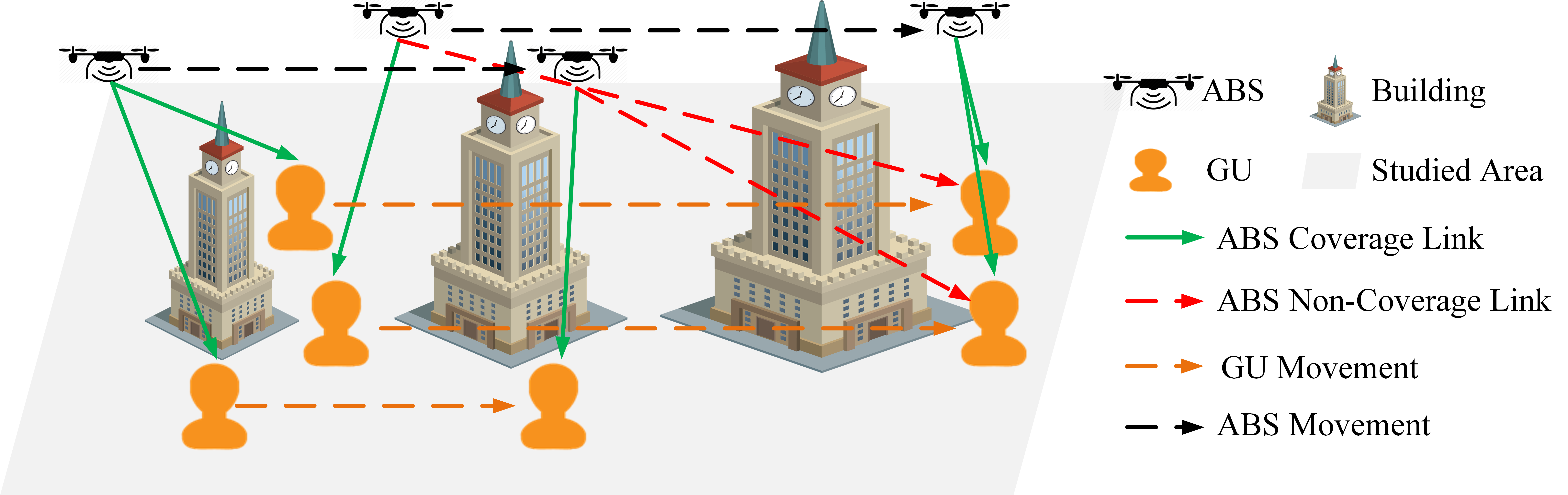}
\caption{Site-specific ABS movement to cover mobile GUs.\vspace{-3ex}}
\label{fig:3D_illustration}
\end{figure}

Consider a UAV-aided communication system with $N$ UAV-mounted ABSs to serve a group of $M$ mobile GUs in a $D_{1}\times D_{2}$ m$^2$ rectangular area with site-specific blockages,  as illustrated in Fig. \ref{fig:3D_illustration}.
For the purpose of exposition, the blockages are exemplified using a collection of $L$ building blocks (BBs), each with a $D_{w}\times D_{w}$ m$^2$ square projection shape and a random height $h_{w}[l]$, $l\in\mathcal{L}\triangleq\{1, \dots, L\}$.
In this work, we concentrate on the access network where ABSs aim to provide data communication coverage for GUs, and assume for simplicity that there exists a backhaul network among ABSs.\footnote{The ABS-ABS channel is more likely to be LoS-dominated which is suitable for establishing a connected backhaul network.}

\subsection{Discretized ABS/GU Plane}\label{SectionPlaneConstruction}
\begin{figure}[!t]
\centering
\includegraphics[width=0.37\textwidth,  trim=10 0 10 0,clip]{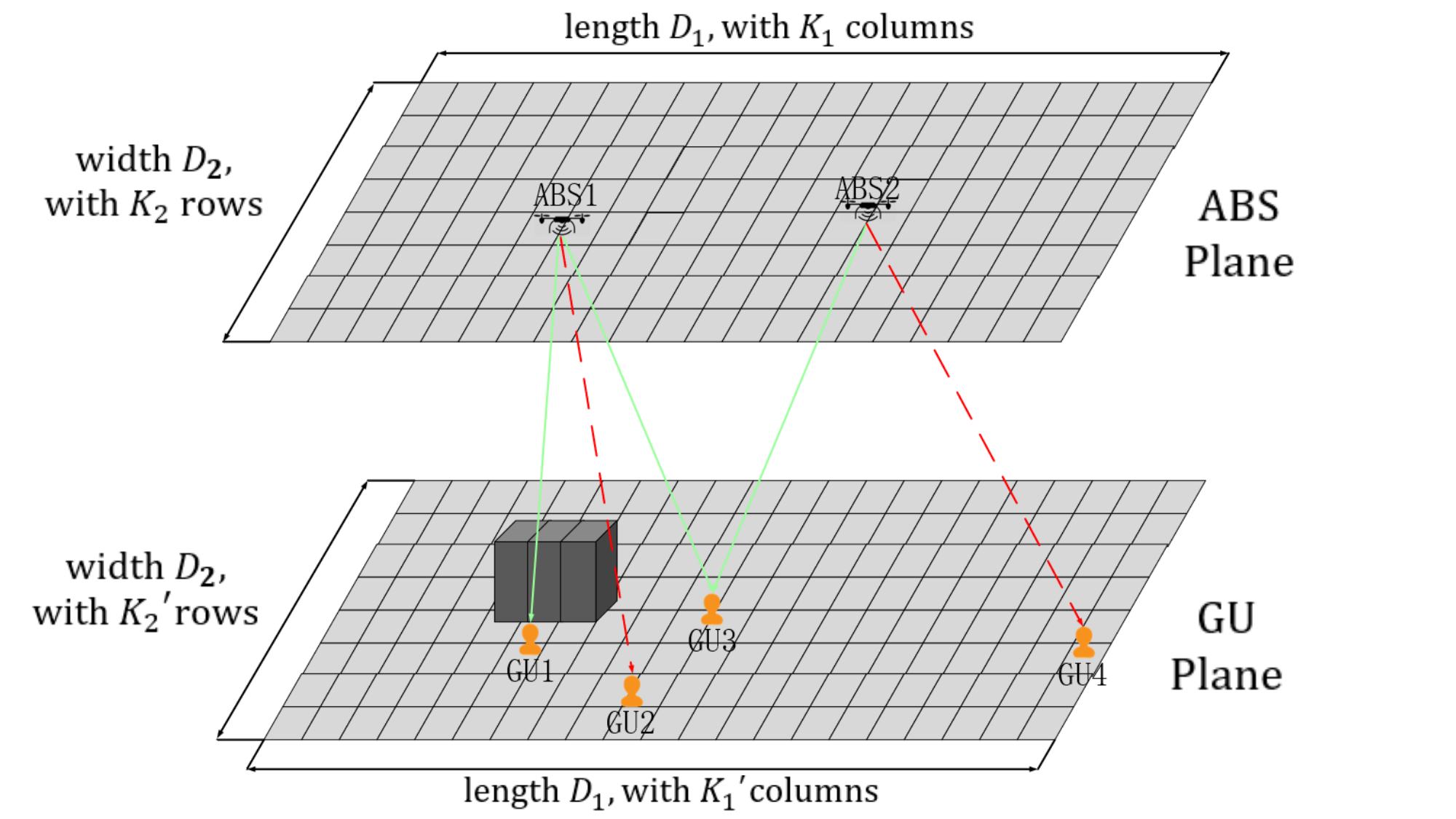}
\caption{Discretized ABS/GU plane.\vspace{-3ex}}
\label{fig:Discretized ABS/GU Plane}
\end{figure}

Fixed altitude at height $H$, for the ABS plane, we consider a $D_{1}\times D_{2}$ area partitioned into $K_{1}\times K_{2}$ grids.
Denote the ABS plane area as $\boldsymbol{\mathcal{A}}$, the grid area as $\mathcal{A}_{ij}$, $\boldsymbol{\mathcal{A}} \triangleq \cup \mathcal{A}_{ij}$, $i = 1,2, \ldots, K_{1}, j = 1,2, \ldots, K_{2}$. 
Each grid has length $\alpha_{1} \triangleq \frac{D_{1}}{K_{1}}$, $\alpha_{2} \triangleq \frac{D_{2}}{K_{2}}$, and grid center location is $\boldsymbol{a}_{i j} \triangleq\left(\left(i-\frac{1}{2}\right) \alpha_{1},\left(j-\frac{1}{2}\right) \alpha_{2}, H\right), i=1,2, \ldots, K_{1}, j=1,2, \ldots, K_{2}$. 
For simplicity, we assume ABSs always locate at the grid center which make up a set $\boldsymbol{A}$, $\boldsymbol{A} \triangleq   \left\{\boldsymbol{a}_{i j} \mid i=1,2, \ldots, K_{1}, j=1,2, \ldots, K_{2}\right\}$. 
For the GU plane, we consider a $D_{1}\times D_{2}$ area partitioned into $K_{1}^{'}\times K_{2}^{'}$ grids on the ground. 
Denote the GU plane area as $\boldsymbol{\mathcal{B}}$, the grid area as $\mathcal{B}_{i^{\prime}j^{\prime}}$, $\boldsymbol{\mathcal{B}} \triangleq \cup \mathcal{B}_{i^{\prime} j^{\prime}}, i^{\prime}=1,2, \ldots, K_{1}{ }^{\prime}, j^{\prime}=1,2, \ldots, K_{2}{ }^{\prime}$.
Each grid has length $\alpha_{1}{ }^{\prime}\triangleq\frac{D_{1}}{K_{1}{ }^{\prime}}, \alpha_{2}{ }^{\prime}\triangleq\frac{D_{2}}{K_{2}{ }^{\prime}}$, 
and grid center location is $\boldsymbol{b}_{i^{\prime} j^{\prime}}\triangleq\left(\left(i^{\prime}-\frac{1}{2}\right) \alpha_{1}{ }^{\prime},\left(j^{\prime}-\frac{1}{2}\right) \alpha_{2}{ }^{\prime}, 0\right),i^{\prime}=1,2, \ldots, K_{1}{ }^{\prime}, j^{\prime}=1,2, \ldots, K_{2}{ }^{\prime}$.
The grid center in GU plane make up a set $\boldsymbol{B}$, $\boldsymbol{B} \triangleq\left\{\boldsymbol{b}_{i^{\prime} j^{\prime}} \mid i^{\prime}=1,2, \ldots, K_{1}{ }^{\prime}, j^{\prime}=1,2, \ldots, K_{2}{ }^{\prime}\right\}$.\footnote{Our discretization operation can be readily extend to 3D UAV movement.}
The discretized ABS and GU planes are as illustrated in Fig. \ref{fig:Discretized ABS/GU Plane}.

\subsection{ABS/GU Mobility Model}\label{SectionMobility}
Consider a typical ABS movement \textit{trial} with a duration of $\Delta T$ s, which is discretized into $I$ equal-length time \textit{steps}, each lasting $\Delta \tau=\Delta T/I$. 
For simplicity, we suppose that ABSs fly at a fixed altitude\footnote{
A statistically optimal ABS altitude can be found based on the channel statistics, in order to maximize its average ground coverage range \cite{2014-Hourani-Optimal}. Real-time three-dimentional (3D) ABS movement is left for future work.} of $H$ m, and GUs move on the ground with a hand-held height of $h_q$ m.
Furthermore, assume that GUs move at a constant pace of $V_q$ m/s but with a random direction at each step. 
We suppose that the positions of GUs at each step are known and communicated through separate control links to a central planning agent.
This planning agent could be located on one of the UAVs or at a ground vehicle station.
Denote $\boldsymbol{p}^{(i)}[n]=(x^{(i)}_{p}[n], y^{(i)}_{p}[n])$ as the horizontal position of ABS $n\in\mathcal{N}\triangleq\{1, \dots, N\}$ at step $i\in \mathcal{I}\triangleq\{1, \dots, I\}$.
Similarly, denote $\boldsymbol{q}^{(i)}[m]=(x^{(i)}_{q}[m], y^{(i)}_{q}[m])$ as the horizontal position of GU $m\in\mathcal{M}\triangleq\{1, \dots, M\}$ at step $i\in\mathcal{I}$.
Denote $\mathcal{P}^{(i)}\triangleq\{\boldsymbol{p}^{(i)}[n]|n\in\mathcal{N}\}$ or $\mathcal{Q}^{(i)}\triangleq\{\boldsymbol{q}^{(i)}[m]|m\in\mathcal{M}\}$ as the location set at step $i\in\mathcal{I}$ for ABSs or GUs, respectively.

Assume each ABS can independently adjust its moving speed as required, subject to a maximum speed constraint of $V^{\max}_p$ m/s. 
Denote $\Vert \cdot \Vert$ as the Euclidean norm. 
Then, the ABS positions in consecutive time steps are restricted by the maximum moving distance, i.e.,
\begin{align}
    \lVert \boldsymbol{p}^{(i)}[n] - \boldsymbol{p}^{(i-1)}[n] \rVert \le V^{\max}_p \cdot \Delta \tau, \forall i\in\mathcal{I}, n\in\mathcal{N}.\label{eq:moving_speed}
\end{align}

Moreover, we focus on the outdoor scenario within a bounded area. Denote $\mathcal{C}\subset \mathcal{A}$ as the region occupied by obstacles. The following constraint is thus imposed, i.e.,
\begin{align}
    \boldsymbol{p}^{(i)}[n]\in\mathcal{A}\setminus \mathcal{C}, \forall i\in\mathcal{I}, n\in\mathcal{N}.\label{eq:area}
\end{align}

\subsection{Site-Specific LoS/NLoS Channel Model}\label{SectionChannelModel}
Consider downlink communication from ABSs to GUs, while our method can also be applied to uplink communication similarly.
To focus on the coverage performance, for simplicity, we assume that the available spectrum is equally partitioned into $M$ orthogonal channels. Each channel is exclusively allocated to an individual GU, thus eliminating intra- or inter-cell interference.
Moreover, suppose that each ABS or GU is equipped with omni-directional antenna of unit gain.\footnote{The case with directional antennas can be similarly considered as in \cite{LyuTWCHotspot}.}
Assume that each ABS transmits with power $P$ Watt (W) to the corresponding served GU, and the receiver noise power is denoted by $\sigma^2$ W.
The SNR received by GU $m$ from ABS $n$ can be expressed as 
\begin{equation}\label{gamma}
\gamma_{m,n}\triangleq g_{m,n}P /\sigma^2,
\end{equation}
where $g_{m,n}\triangleq \bar g_{m,n} \xi_{m,n}$ is instantaneous channel power gain, with $\bar g_{m,n}$ representing the average channel power and $\xi_{m,n}$ accounting for small scale fading with unit average power.


Due to site-specific blockages, the ABS-GU channel could be in either LoS or NLoS condition depending on whether there are obstacles in between.
Therefore, the average channel power gain between GU $m$ and ABS $n$ can be expressed as
\begin{align}\label{probLOS}
\bar g_{m,n}\triangleq
\begin{cases}
\bar g_{\textrm{L}}(\boldsymbol{p}[n], \boldsymbol{q}[m]), & \quad \textrm{no obstacles in between;}\\
\bar g_{\textrm{NL}}(\boldsymbol{p}[n], \boldsymbol{q}[m]), & \quad \textrm{otherwise,}
\end{cases}
\end{align}%
where $\bar g_{\textrm{L}}$ and $\bar g_{\textrm{NL}}$ denote the average channel power gains of the LoS and NLoS channels, respectively.\footnote{As a preliminary study, we adopt the urban macro formulas in 3GPP \cite{3GPP} as the underlying path-loss model in our simulations.}
Regarding small-scale fading, for the LoS case, consider the angle-dependent Rician fading channel with factor $K_{m,n}$ given by \cite{10373821}
\begin{equation}\label{RicianK}
    K_{m,n} = A_{1}\textrm{exp}(A_{2}\theta_{m,n}),
\end{equation}
where $A_{1}$ and $A_{2}$ are constant coefficients, and $\theta_{m,n}\triangleq \arctan\frac{h_p-h_q}{\lVert \boldsymbol{p}[n] - \boldsymbol{q}[m] \rVert}$ is the elevation angle of ABS $n$ as seen by GU $m$.
Then we have $K_{\textrm{min}} \leq K \leq K_{\textrm{max}}$, where $K_{\textrm{min}} = A_{1}$ and $K_{\textrm{max}} = A_{1}e^{A_{2}\pi/2}$. On the other hand, for the NLoS case, Rayleigh fading is considered which is a special case of Rician fading with $K_{m,n}=0$.\footnote{
Note that the channel model in \eqref{probLOS} and \eqref{RicianK} only serves as the underlying ground truth model used in the simulation studies. Our proposed scheme is based on GCM which could be obtained by on-site connectivity measurements\cite{YangUAV} or other RM construction methods\cite{2019-Lyu-RadioMap}\cite{2023-liu-UAV-Aided}\cite{levie2021radiounet}\cite{zhen2022radio}.}

Due to small scale fading, the instantaneously received SNR might fall below a certain required level $\bar\gamma$ and cause communication outage, with outage probability $P_{\text{out},m}\triangleq \textrm{Pr}\{\gamma_{m,n}<\bar\gamma\}$. 
As a result, a GU $m$ is considered covered by ABS $n$, if the outage probability is below a certain threshold $\eta$.

\subsection{Global Connectivity Map and Problem Formulation}\label{SectionGCM}
In this section, we construct GCM to formulate the considered problem into a BILP problem. Define $u \triangleq(i-1) K_{2}+j$, $i=1,2, \ldots, K_{1}$, $ j=1,2, \ldots, K_{2}$ as the flattened index of ABS grid $(i, j)$ on the ABS plane, and $v \triangleq\left(i^{\prime}-1\right) K_{2}{ }^{\prime}+j^{\prime}$, $i^{\prime}=1,2, \ldots, K_{1}{ }^{\prime}$, $j^{\prime}=1,2, \ldots, K_{2}{ }^{\prime}$ as the flattened index of GU grid $(i^{\prime}, j^{\prime})$ on the GU plane.
With slight abuse of notations, we use $u$ and $(i,j)$, $v$ and $(i^{\prime},j^{\prime})$ interchangeably (\text{e.g.,} $\boldsymbol{a}_{u}=\boldsymbol{a}_{ij}$, $\boldsymbol{b}_{v}=\boldsymbol{b}_{i^{\prime} j^{\prime}}$).
For a given grid  $u$ on the ABS plane and grid $v$ on the GU plane, define $z_{u v}$ as the connectivity indicator which is given by
\begin{align}\label{zuv}
    z_{u v}  = \left\{\begin{array}{l}
        1, \text { if } P_{\text{out}}\left(\boldsymbol{a}_{u}, \boldsymbol{b}_{v}\right) < \eta ;\\
        0, \text { otherwise, }
        \end{array}\right.
\end{align}
where  $\eta$  is a predefined threshold. Here we use the grid center $b_{v}$ to represent any given GU located in the grid $v \in V$. 
Such approximation significantly simplifies the problem at the cost of quantization error, which will be evaluated in Section \ref{gridlength}.

The resulted GCM is a binary matrix $\boldsymbol{Z}$,  with the  $(u, v)$-th element given by
$[\boldsymbol{Z}]_{u v} \triangleq z_{u v}, u\in U\triangleq\{1,2, \ldots, K_{1} K_{2}\}, v\in V\triangleq\{1,2, \ldots, K_{1}{ }^{\prime} K_{2}{ }^{\prime}\} $.
According to GCM, we can then define a coverage indicator $C_{v}^{(i)}$ for GU on grid $v$ in step $i$ as
\begin{equation}\label{Cv}
 C_{v}^{(i)}  \triangleq \min\left\{\sum\nolimits_{
 \substack{u \in U, v \in V}}
 a_{u}^{(i)}b_{v}^{(i)}{z_{u v}},1\right\}, 
\end{equation}%
where the binary element $a_{u}, b_{v} \in \left\{0, 1\right\}$ indicate whether there exist ABS/GU on the corresponding grid. The variable $a_{u}$ is constrained by the total number of available ABSs, i.e.,
\begin{align}\label{au_sum}
    \sum\nolimits_{u\in U}a_{u} = N. 
\end{align}

Considering the constraints \eqref{eq:moving_speed} and \eqref{eq:area}, we can obtain the feasible ABS grid index set $U_{n}^{'}$ which is within a circle with ABS $n$  as the center and the maximum moving distance as the radius. Denote $U^{'}$ as the union of $U_{n}^{'}, n\in \mathcal{N}$. We deploy at least one ABS in each feasible set,
\begin{align}\label{au_deploy}
\sum\nolimits_{u \in U_{n}^{'}} a_{u}\geq 1,  n \in \mathcal{N}.
\end{align}

Then, we decompose the nonlinear formula \eqref{Cv} into three equivalent linear formulas to effectively reduce the computational complexity of the problem, i.e.,

\begin{align}\label{decompose2}
C_{v}^{(i)} \geq a_{u}^{(i)}b_{v}^{(i)}z_{u v}, u \in U^{'}, v \in V,
\end{align}
\begin{align}\label{decompose3}
C_{v}^{(i)} \leq \sum\nolimits_{u \in U^{'}} a_{u}^{(i)} b_{v}^{(i)} z_{u v}, v \in V,
\end{align}
\begin{align}\label{decompose1}
    C_{v}^{(i)}\in\left\{0,1\right\},v \in V.
\end{align}

The coverage rate at step $i$ is determined as follows
\begin{equation}\label{CR}
    \lambda^{(i)}\triangleq \frac{1}{M}\sum\nolimits_{v \in V} C_{v}^{(i)}.
\end{equation}

Our objective is to maximize the average coverage rate (ACR) $\bar\lambda$ over the entire trial through multi-ABS movement optimization, as given by
\begin{equation*}
    \begin{aligned}
        \text{(P1):}~\max~&\bar\lambda\triangleq\frac{1}{I}\sum\nolimits_{i \in\mathcal{I}}\lambda^{(i)}\\
        \text{s.t.} ~&  \eqref{zuv},\eqref{au_sum}, \eqref{au_deploy},  \eqref{decompose2},\eqref{decompose3},\eqref{decompose1}~\text{and}~\eqref{CR}.
    \end{aligned}
\end{equation*}%

The problem (P1) is a BILP problem, which is still an NP-hard problem according to \cite{luenberger1984linear}. The BILP problem can be traditionally solved  by utilizing open source solver (e.g., SCIP \cite{BestuzhevaEtal2021ZR}).
However, with the expansion of the environment map, the variables and constraints in (P1) increase dramatically, leading to significant time consumption when solving the problem. Since (P1) requires to find a feasible ABS location set within a short time period,  a method which can solve it rapidly and effectively is desired. 
To tackle the above problem, we introduce the novel fast online algorithm \cite{li2020simple} with tailored modifications to solve (P1) efficiently. 

\section{Fast Oniline Algorithm for ABS Movement}\label{sec:proposed_algorithm}

In this section, we introduce a three-level time hierarchy to partition the ABS movement problem into ABS placement sub-problems, which are then solved by the fast online algorithm.

\subsection{Three-Level Time Hierarchy}

\begin{figure}[!t]
\centering
\includegraphics[width=0.38\textwidth,  trim=50 0 50 0,clip]{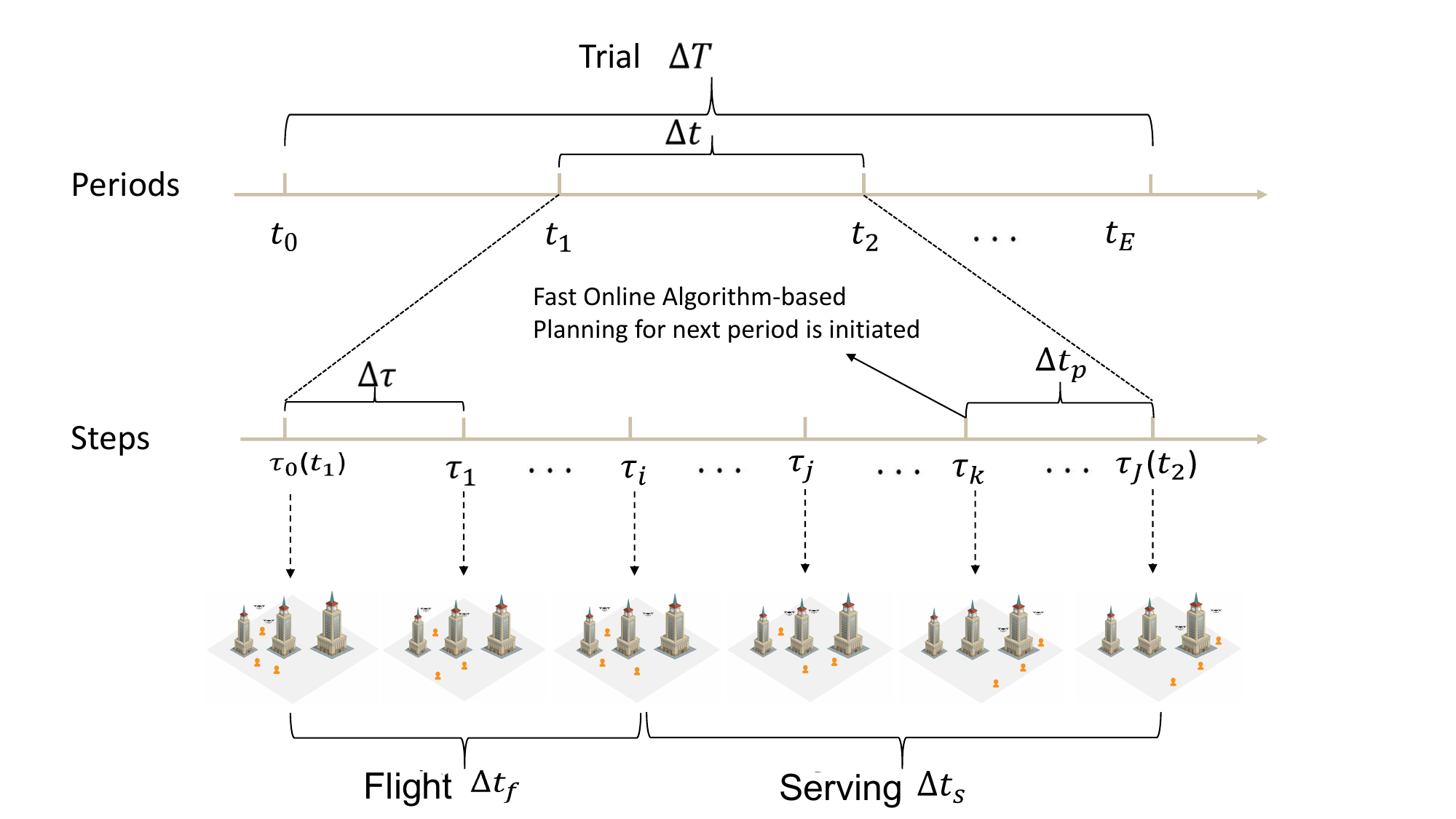}
\caption{Trial-Period-Step time hierarchy.\vspace{-3ex}}
\label{fig:hierarchy}
\end{figure}

One trial for problem (P1) is equally divided into $E$ \textit{periods}, where each period has $J\triangleq I/E$ steps and thus lasts for $\Delta t\triangleq\Delta T/E= J\cdot\Delta \tau$ s, as illustrated in Fig. \ref{fig:hierarchy}. 
As a preliminary study, we focus on scenarios characterized by low/moderate GU mobility, and assume that the distribution of GUs between two consecutive periods exhibit only minor variations.
Each period consists of two non-overlapping phases, i.e., flight and serving, each with a duration of $\Delta t_{f}$ s and $\Delta t_{s}$ s, respectively, with $\Delta t_{f}+\Delta t_{s}=\Delta t$. 
For a target period (e.g., $t_1\sim t_2$ in Fig. \ref{fig:hierarchy}), a preliminary planning procedure spanning $\Delta t_p$ ($\Delta t_p\leq \Delta t$) is initiated beforehand, 
which utilizes the most recently reported positions of GUs as input,
and uses the fast online algorithm to output the candidate ABS placement solutions.
Each ABS adjusts its moving speed to reach the destination in the flight period and hover until the end of the serving period. 
Next, we focus on our proposed fast online algorithm.

\subsection{Fast Online Algorithm}\label{MHFO algorithm}
Due to the NP-hard nature of the problem (P1), using traditional open source solver (SCIP) might impose high computational complexity to solve such ABS placement problem as the problem size increases.
In \cite{li2020simple}, the author proposes a fast online algorithm which is used to solve a class of BILP problems rapidly, whereby the obtained solution achieves the close-to-optimal performance.  
It essentially conducts  (one-pass) projected stochastic subgradient descent within the dual space.
Here we use the fast online algorithm \cite{li2020simple} with tailored modifications to solve our BILP problem. Firstly, we rewrite the constraints \eqref{au_sum} $\sim$ \eqref{decompose1} into matrix and vector form as
\begin{equation*}
\scalebox{0.73}{
\resizebox{1\linewidth}{!}{
  $\begin{bmatrix}
    0 & \ldots & 0 & 1 & \ldots & 1 \\
    0 & \ldots & 0 & \ldots & \boldsymbol{-1}_{U_1'} & \ldots \\
    \vdots & \ddots & \vdots & \vdots & \ddots & \vdots \\
    0 & \ldots & 0 & \ldots & \boldsymbol{-1}_{U_N'} & \ldots \\
    -1 & \ldots & 0 & b_{1}z_{11} & \ldots & 0 \\
    \vdots & \ddots & \vdots & \vdots & \ddots & \vdots \\
    0 & \ldots & -1 & 0 & \ldots & b_{V}z_{U'V} \\
    1 & \ldots & 0 & -b_{1}z_{11} & \ldots & -b_{1}z_{U'1} \\
    \vdots & \ddots & \vdots & \vdots & \ddots & \vdots \\
    0 & \ldots & 1 & -b_{V}z_{1V} & \ldots & -b_{1}z_{U'V} \\
  \end{bmatrix}
  \begin{bmatrix}
    C_{1} \\
    \vdots \\
    C_{V} \\
    a_{1} \\
    \vdots \\
    a_{U'}
  \end{bmatrix}
  \leq
  \begin{bmatrix}
    N \\
    -1 \\
    \vdots \\
    -1 \\
    0 \\
    \vdots \\
    0 \\
    0 \\
    \vdots \\
    0
  \end{bmatrix}$
}}
\end{equation*}
where $\boldsymbol{1}_{U_n'}\in \mathbb{R}^{(U^{'})}$ is an elementwise indicator function for the feasible ABS grid index set $U_{n}^{'}$ of each ABS $n$. Denote $\boldsymbol{E}\in\mathbb{R}^{(1 +N+ U^{'}V + V)\times(V + U^{'} )}$ as the coefficient constraint matrix, 
$\boldsymbol{x}\in \mathbb{R}^{(V + U^{'})}$ as the variable vector, and
$\boldsymbol{l}\in \mathbb{R}^{(1 + N+U^{'}V + V)}$ as the right-hand side constant vector.
Note that the superscript $(i)$ is omitted here for simplicity, and we focus on maximizing the number of covered GUs for a given ABS placement subproblem, as given by
\begin{equation*}
\small
    \begin{aligned}
        \text{(P2):}~\max~&\sum\nolimits_{\substack{v\in V}}C_{v}\triangleq\boldsymbol{r}^{T}\boldsymbol{x}\\
        \text{s.t.} ~& \boldsymbol{E}\boldsymbol{x}\leq\boldsymbol{l},
        x_{k}\in\left\{0,1\right\},k \in (V + U^{'}),
    \end{aligned}
\end{equation*}%
where $\boldsymbol{r} = (1,\ldots,1, 0,\ldots,0)\in\mathbb{R}^{(V + U^{'})}$ denote the coefficient vector of the objective function. The problem (P2) is a BILP problem involving integer variables. We first apply linear relaxation to the above problem (P2), as given by
\begin{equation*}
    \begin{aligned}\label{plp}
        \text{(P3):}~\max~& \boldsymbol{r}^{T}\boldsymbol{x}\\
        \text{s.t.} ~& \boldsymbol{E}\boldsymbol{x}\leq\boldsymbol{l},
        \boldsymbol{0}\leq\boldsymbol{x}\leq\boldsymbol{1}.
    \end{aligned}
\end{equation*}%
According to \cite{luenberger1984linear}, the dual linear programming (DLP)  problem of (P3) is given by
\begin{equation*}
    \begin{aligned}\label{dlp}
        \text{(P4):}~\min~& \boldsymbol{l}^{T}\boldsymbol{y} + \boldsymbol{1}^{T}\boldsymbol{s}\\
        \text{s.t.} ~& \boldsymbol{E}^{T}\boldsymbol{y}+\boldsymbol{s}\geq\boldsymbol{r},
        \boldsymbol{y}\geq\boldsymbol{0},\boldsymbol{s}\geq\boldsymbol{0},
    \end{aligned}
\end{equation*}%
where dual decision variables are $\boldsymbol{y} \in \mathbb{R}^{(1 + N+U^{'}V + V)}$, $\boldsymbol{s} \in \mathbb{R}^{(V + U^{'} )}$.
Denote the optimal solution to (P3) and (P4) as $\boldsymbol{x}^{*}$, $\boldsymbol{y}^{*}$ and $\boldsymbol{s}^{*}$. 
Based on complementary slackness condition\cite{luenberger1984linear} 
\begin{align}\label{complementary}
x_{k}^{*}  = \left\{\begin{array}{ll}
1, & r_{k}>\boldsymbol{e}_{k}^{T} \boldsymbol{y}^{*}; \\
0, & r_{k}<\boldsymbol{e}_{k}^{T} \boldsymbol{y}^{*},
\end{array}\right.
\end{align}
where $\boldsymbol{e}_{k}$ denotes $k$-th column of the matrix $\boldsymbol{E}$. Plug the constraints in (P4) to its objective function to remove the dual decision variables $\boldsymbol{s}$. The reformulated problem is given by
\begin{align}\label{new_obj_function}
~\min~& \boldsymbol{l}^{T}\boldsymbol{y} + \boldsymbol{1}^{T}(\boldsymbol{r} - \boldsymbol{E}^{T}\boldsymbol{y})\\
\text{s.t.}~&\boldsymbol{y}\geq\boldsymbol{0}\notag,
\end{align}
For more refined algorithm derivation, we divide the matrix operations into certain vector operations and denote $\boldsymbol{l} = (V + U^{'})\boldsymbol{d}$. Extract the constant coefficient $(V + U^{'})$, and we obtain an equivalent form of (P4) that only involves decision variable $\boldsymbol{y}$, as given by 
\begin{align}\label{equivalent form}
\min _{\boldsymbol{y} \geq \mathbf{0}}~& f(\boldsymbol{y})  = \boldsymbol{d}^{T} \boldsymbol{y}+\frac{1}{V + U^{'}} \sum\nolimits_{k  = 1}\nolimits^{V + U^{'}}\left(r_{k}-\boldsymbol{e}_{k}^{T} \boldsymbol{y}\right)^{+}
\end{align}
where $(\cdot)^{+}$ denotes the positive part function.
The transition from $\boldsymbol{y}_{t}$ to $\boldsymbol{y}_{t+1}$ can be viewed as implementation of the projected stochastic subgradient descent technique for optimizing problem \eqref{equivalent form}.
At each iteration $t$, it updates the vector with the latest observation and performs a projection onto the non-negative orthant to maintain dual feasibility.
Concretely, the subgradient of the $t$-th term in \eqref{equivalent form} evaluated at $\boldsymbol{y}_{t}$ is as follows,
\begin{align}\label{sbugredient update}
&\left.\partial_{\boldsymbol{y}}\left(\boldsymbol{d}^{T} \boldsymbol{y}+\left(r_{t}-\boldsymbol{e}_{t}^{T} \boldsymbol{y}\right)^{+}\right)\right|_{\boldsymbol{y}= \boldsymbol{y}_{t}}\notag\\
&= \boldsymbol{d}-\left.\boldsymbol{e}_{t} \mathbb{I}\left(r_{t}>\boldsymbol{e}_{t}^{T} \boldsymbol{y}\right)\right|_{\boldsymbol{y}  = \boldsymbol{y}_{t}}= \boldsymbol{d}-\boldsymbol{e}_{t} x_{t},
\end{align}
where $\mathbb{I}(\cdot)$  denotes the indicator function.
We denote $\Xi(\mathcal{P})$ as the function to calculate the currently covered GUs, $\vee$ as the elementwise maximum operator, and $\alpha$ as the step size. According to \cite{gao2023solving}, we also apply the duplication factor $K$ to increase the covered GU number.
The pseudo code of the fast online algorithm is described in Algorithm \ref{alg:online}.

The algorithm create the first loop by applying the duplication factor $K$ (Line 1).
Since the number of variable $C_{v}$ is connected to the number of GU $M$, and GU location set is known in advance, we focus on the grid that located GU. The algorithm initializes a random permutation $q$ which is a random variable index array ranging from 0 to $M + U^{'} - 1$, whereby improving the randomness of the algorithm. Each element in array $q$ indicates the variable index (Line 2). 
Then, from the complementary slackness condition \eqref{complementary}, we utilize constraint matrix $\boldsymbol{E}$ and dual decision variables $\boldsymbol{y}$ to find feasible ABS locations in the second loop (Line 3$\sim$4).
Due to the implementation of the projected stochastic subgradient descent technique, we update dual decision variable $\boldsymbol{y}$ based on \eqref{sbugredient update} (Line 5).
Finally, obtain $\mathcal{P}_{\text{tmp}}$ which is the ABS placement location set of current loop based on $\boldsymbol{x}$. After $K$ loops, we find the algorithm-believed best ABS location set $\mathcal{P}^{*}$ (Line 8$\sim$10).

\subsection{Optimality and Complexity Analysis}
The worst time complexity of this algorithm is $\mathcal{O}(K(M + K_{1}K_{2}))$. By considering the ABS movement range constraint, we effectively reduce the time complexity to $\mathcal{O}(K(M + U^{'}))$, whereby $U^{'}$ is significantly smaller than the global range $K_{1}K_{2}$.
Based on \cite{li2020simple}, we denote the optimal objective values of the problem P2 and P3, as $Q_{n}^{*}$ and $R_{n}^{*}$, respectively. The objective value obtained by the algorithm
output is denoted as $R_{n}$. The quantity of interest is the optimality gap $Q_{n}^{*} - R_{n}$, which has an upper bound $Q_{n}^{*} - R_{n} \leq R_{n}^{*} - R_{n}$. The expected optimality gap is defined as $\mathbb{E} [R_{n}^{*} - R_{n}]$. According to Theorem 1 in \cite{li2020simple}, the expected optimality gap is $\mathbb{E} [R_{n}^{*} - R_{n}] \leq (1 + N + U^{'}V + V)(\overline{e} + \overline{d})^{2}\sqrt{V + U^{'}}/\sqrt{K}$, where $\| \boldsymbol{e}_{k} \|_{\infty} \leq \overline{e}, k = 1, \ldots,V + U^{'}$, $\| \boldsymbol{d}_{i} \|_{\infty} \leq \overline{d}, i = 1, \ldots,1 + N + U^{'}V + V,$ and $K$ is the duplication factor. 
\begin{algorithm}[!t]\caption{Fast Online Algorithm for ABS Placement}\label{alg:online}
	\begin{small}
	\textbf{Input:} Original ABS set $\mathcal{P}$, GU set $\mathcal{Q}$, feasible ABS placement index set $U^{'}$ and duplication factor $K$.\\
	\textbf{Output:}  Best ABS location set  $\mathcal{P}^{*}$.
		
    \begin{algorithmic}[1]
        \STATE \textbf{for} $i$ \textbf{in} $K$ \textbf{do}
		\STATE \quad Initialize random permutation $q$.
		\STATE \quad \textbf{for} $t$ \textbf{in} $q$ \textbf{do}
        \STATE \quad\quad  $x_{t} = \left\{\begin{array}{ll}
1, & r_{t}>\boldsymbol{e}_{t}^{T} \boldsymbol{y}_{t} \\
0, & r_{t}<\boldsymbol{e}_{t}^{T} \boldsymbol{y}_{t}
\end{array}\right.$
        \STATE \quad\quad $\boldsymbol{y}_{t+1}=\boldsymbol{y}_{t}+\alpha\left(\boldsymbol{e}_{t} x_{t}-\boldsymbol{d}\right), \boldsymbol{y}_{t+1}=\boldsymbol{y}_{t+1} \vee \boldsymbol{0}$
        \STATE \quad \textbf{end for}
        \STATE \quad Obtain $\mathcal{P}_{\text{tmp}}$ based on $\boldsymbol{x}$.
        \STATE \quad \textbf{if} $\Xi$($\mathcal{P}_{\text{tmp}}$) $>$ $\Xi$($\mathcal{P}^{*}$) :
        \STATE \quad\quad $\mathcal{P}^{*}\leftarrow\mathcal{P}_{\text{tmp}}$
        \STATE \textbf{end for}
    \end{algorithmic}
	\end{small}
\end{algorithm}

\section{Numerical Results}
\label{sec:simulation_result}
This section presents comprehensive simulation results showcasing the effectiveness of our proposed algorithm.
Three benchmarks are provided. The upper bound performance is obtained by the open-source solver (SCIP) \cite{BestuzhevaEtal2021ZR}.
In addition, we implement one of the state-of-the-art DRL methods (TD3) \cite{10373821}\cite{2020-Lyu-Codesign}, where the learning rates of the actor and critic are both 0.0003, and the discount factor is 0.995, with more detailed hyperparameters in \cite{10373821}.
Finally, we introduce the K-means initiated evolutionary algorithm (EA) which initially employs the K-means algorithm to generate initial ABS locations\cite{2016-Galkin-Deployment}, and then utilizes the EA-based mutation strategy within the specified mutation range to generate mutated ABS sets during the planning period\cite{10373821}. Each mutation set is subject to the constraints \eqref{eq:moving_speed} and \eqref{eq:area}. 
In $K_{m}$ rounds of mutation, we select the ABS set with the highest CR as the next positions.
The performance metrics include the ACR, the average planning time ($\overline{t}$) of whole trial, and the step-wise CR.
The default parameters are listed below:
$D_{1}=1000$ m, $D_{2}=1000$ m, $K_{1}=K_{1}^{'}=40$, $K_{2}=K_{2}^{'}=40$, $h_w[l] \sim \textrm{Uniform}(30,89)$ m, $L=300$, $H=90$ m, $h_{q}=1$ m, $V^{\max}_p=30$ m/s, $V_q=2$ m/s, $K=3$, $K_{m}=3000$, $\Delta T=200$ s, $\Delta t=20$ s, $\Delta t_f=10$ s, $\Delta t_s=10$ s, $\Delta t_p=5$ s, $\Delta \tau=1$ s, $P=5$ dBm, $\sigma^2=-112$ dBm, $K_{\textrm{min}}=0$ dB, $K_{\textrm{max}}=30$ dB, $\bar{\gamma}=3$ dB, and $\eta=0.1$.

\subsection{ABS Trajectory Visualization}
\begin{figure}[!t]
    \centering
        \begin{subfigure}[b]{0.46\linewidth}
        \centering
        \includegraphics[width=\linewidth,  trim=10 70 0 70,clip]{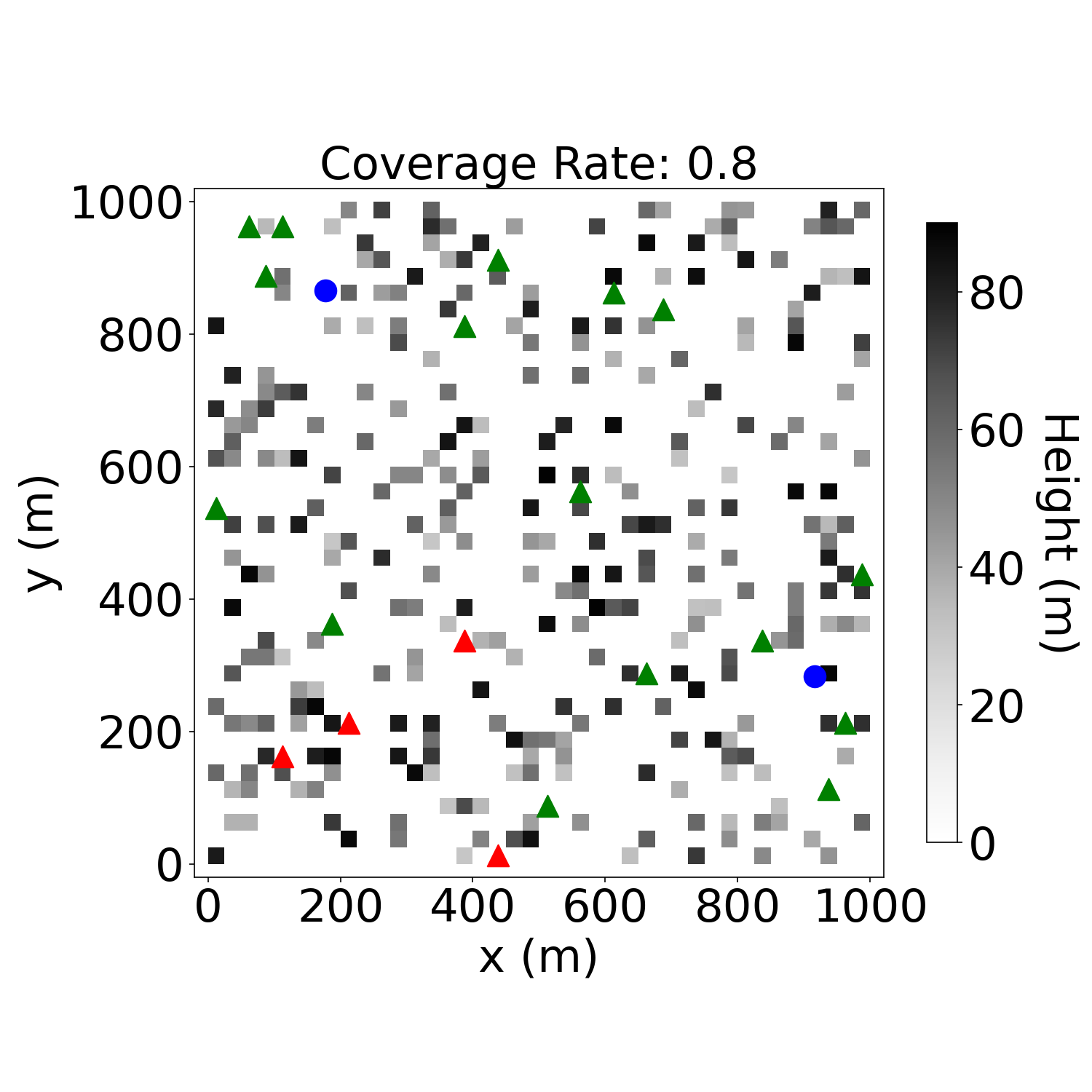}
        \caption{initial locations \vspace{-1ex}}
        \label{fig:step1}
    \end{subfigure}
    \begin{subfigure}[b]{0.46\linewidth}
        \centering
        \includegraphics[width=\linewidth,  trim=10 150 0 70,clip]{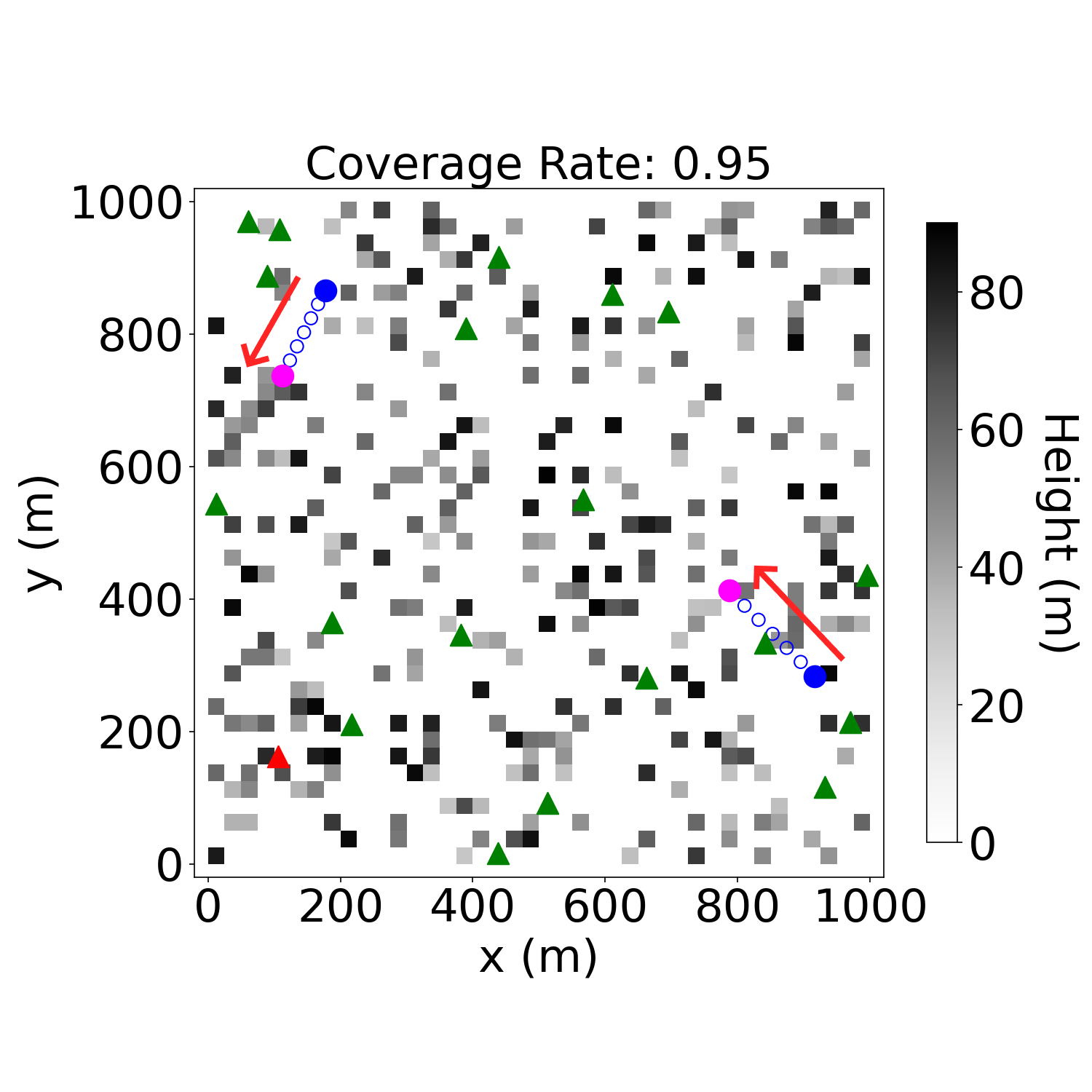}
        \caption{flight trajectory \vspace{-1ex}}
        \label{fig:step2}
    \end{subfigure}
    \caption{Illustration of ABS flight trajectory during consecutive steps ($N=2$, $M=20$). The squares represent the generated BBs, whose color darkness indicates their heights. The initial ABSs is denoted by blue points, and the final ABSs is denoted by magenta points. The hollow points represent the positions of ABSs during the flight period. Triangles symbolize GUs, with green and red coloring indicating covered and uncovered GUs, respectively. \vspace{-1ex}}
    \label{fig:TRAJECTORY}
\end{figure}

For illustration, we first select several consecutive steps to visualize the ABS flight trajectory and GU coverage status. The initial locations and flight trajectory are shown in Fig. \ref{fig:TRAJECTORY}.
For the considered example, the coverage rate is improved from 0.8 to 0.95 by utilizing our proposed algorithm.

\subsection{Grid Length Sensitivity and Quantization Error}\label{gridlength}
In this subsection, we change the grid length to analyze the quantization error of the approximation in Section \ref{SectionGCM}. In the simplified situation, we use the grid center $\boldsymbol{b}_{v}$ to represent any given GU located in the grid $v$, while in the actual situation, we use the real locations of any given GU. The results of different grid lengths and quantization errors are shown in Table \ref{tab:grid}.

It can be seen from the Table \ref{tab:grid} that the approximation produces a certain quantization error while simplifying the problem, but when the grid length decreases, the quantization error between the simplified situation and the actual situation shows a gradually decreasing trend. Moreover, due to the decrease of grid length, more grids can be chosen for the ABS placement, thus improving the average coverage rate. Therefore, there exists a general trade-off between the CR performance, quantization error and the algorithm complexity, whereby the grid length in a given area needs to be chosen carefully depending on the application requirements.
Finally, note that the quantization error can be further reduced by employing spatial interpolation techniques such as kriging.
\begin{table}[htbp]\small
	\centering
	\caption{Grid length sensitivity and quantization error ($N=5$, $M=100$, trials = $20$).}
	\label{tab:grid}
	\resizebox{0.38\textwidth}{!}{%
		\begin{tabular}{|c|c|c|c|}
			\hline
			\textbf{Grid Length(m)}  & \textbf{12.5} & \textbf{25} & \textbf{50}  \\ \hline
			\textbf{Simplified Situation(ACR)} & 0.942           & 0.917           & 0.891           \\ \hline
			\textbf{Actual Situation(ACR)}  & 0.929           & 0.891            &0.855                       \\ \hline
			\textbf{Quantization Error}  & 0.013           & 0.026            &0.036                       \\ \hline                
		\end{tabular}%
	}
\end{table}
\subsection{Step-wise Coverage Performance}
\begin{figure}[!t]
	\centering
	\includegraphics[width=0.36\textwidth,  trim=0 0 0 0,clip]{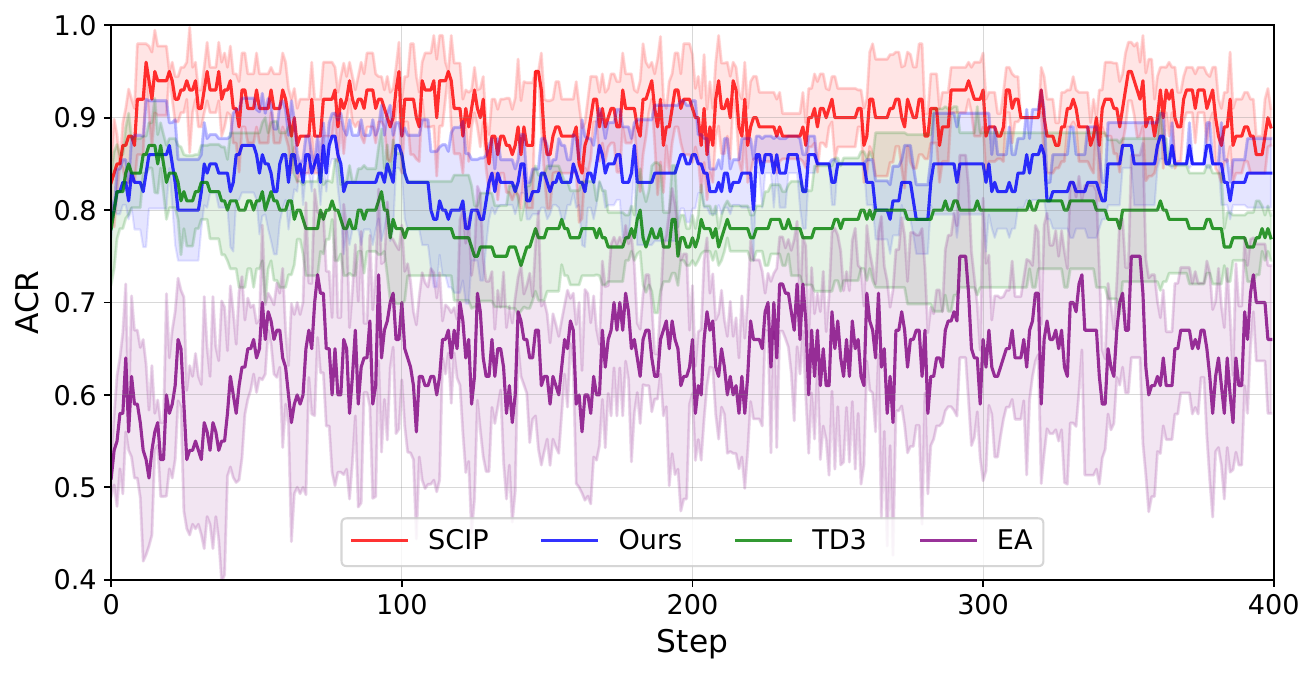}
	\caption[]{Average coverage rate of the four different algorithms with $400$ steps over 5 different trials($N=2$, $M=20$).}\vspace{-3ex}
	\label{fig:plot}
\end{figure}

\begin{figure}[!t]
	\centering
	\includegraphics[width=0.36\textwidth,  trim=0 0 0 0,clip]{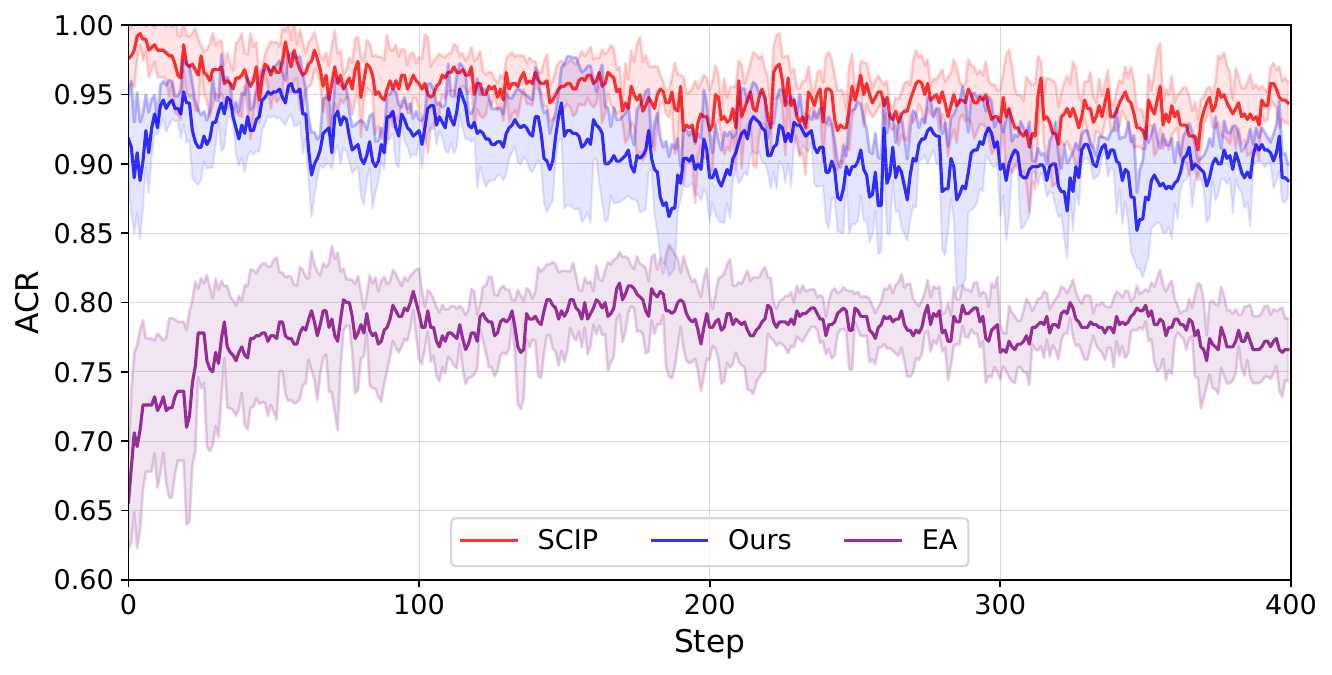}
	\caption[]{Average coverage rate of the three different algorithms with $400$ steps over 5 different trials($N=5$, $M=100$).}\vspace{-3ex}
	\label{fig:plot1}
\end{figure}

In this section, we consider two dynamic scenarios with moving GUs 
and evaluate the performance (ACR, $\overline{t}$). 
Note that the TD3 algorithm fails to converge when faced with larger-scale problems.
The results are shown in Table \ref{tab:performance}, Fig. \ref{fig:plot} and Fig. \ref{fig:plot1}.
Our algorithms exhibit superior performance compared to the TD3 and EA methods in terms of average CR in both two cases, 
meanwhile, our algorithm significantly reduces average planning time compared to using SCIP directly, due to the implementation of the fast online approach that swiftly produces ABS location set.
Owing to the real-time limitations inherent in our scenario, it should be highlighted that the outcomes delivered by SCIP represent a theoretical best case that may not be feasible in practical applications.


%

\subsection{Accommodation To Complex/Dynamic Scenarios}
\begin{figure}[!t]
	\centering
	\includegraphics[width=0.36\textwidth, trim=0 0 0 0,clip]{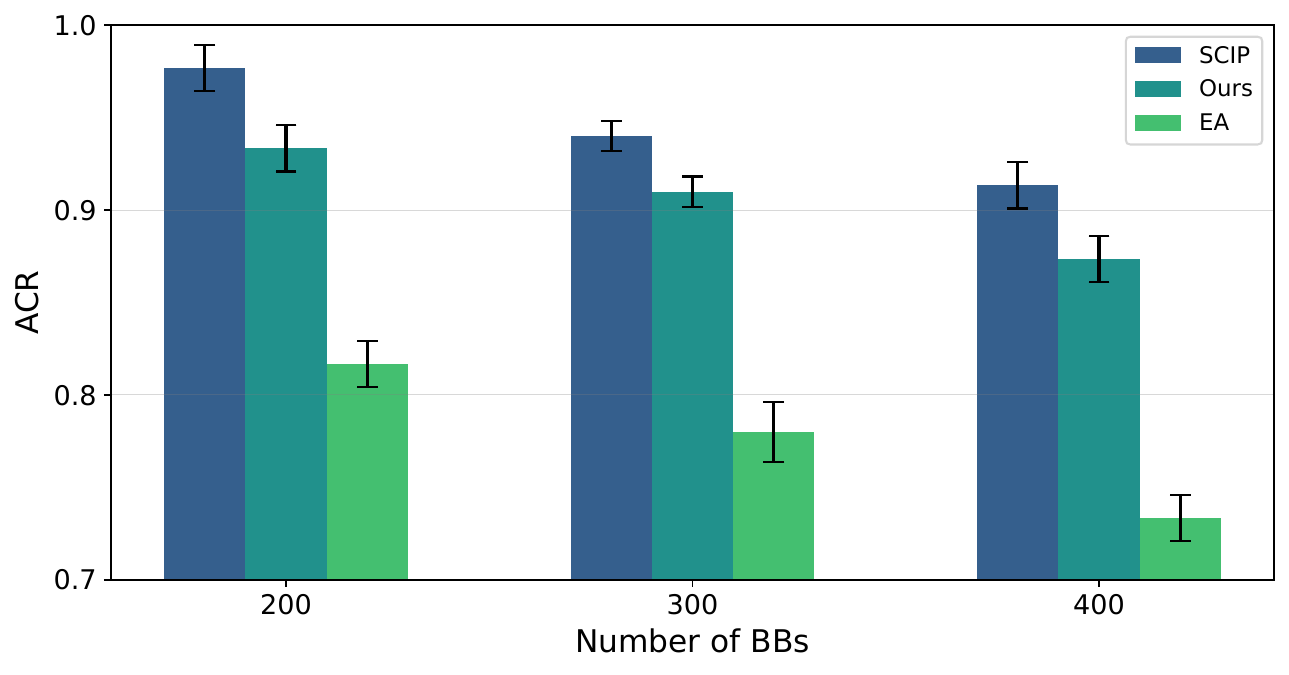}
	\caption{Average coverage rate of  different number of BBs ($N=5$, $M=100$). The black error bars indicate the standard deviations.}\vspace{-3ex}
	\label{fig:plot_bb}
\end{figure}

We also compare the performance under different site-specific environments (e.g., with different number of BBs), as shown in Fig. \ref{fig:plot_bb}.
Notably, our proposed algorithm consistently outperforms the EA method, and the disparity between ours and SCIP remains relatively stable even as the number of BBs increases.
In addition, we also change GU speeds (2$ \sim $10 m/s) to observe ACR performance of different algorithms.
It is observed that our proposed algorithm still achieves a high ACR under moderate GU speeds, and consistently outperforms EA methods. The detailed results are omitted for brevity.
\begin{table}[!t]\small
	\centering
	\caption{The average performance in different scenarios.}
	\label{tab:performance}
	\resizebox{0.36\textwidth}{!}{%
		\begin{tabular}{|c|c|c|c|c|}
			\hline
			\textbf{Algorithms}  & \textbf{SCIP} & \textbf{Ours} & \textbf{TD3} &\textbf{EA}  \\ \hline
			\textbf{ACR($N=2$, $M=20$)} & 0.90           & 0.84           & 0.79       &0.64    \\ \hline
			\textbf{$\overline{t}$(s)($N=2$, $M=20$)}  & 1.47           & 0.92       &0.96  &1.18             \\ \hline
            \textbf{ACR($N=5$, $M=100$)} & 0.95           & 0.91           &    $ - $     & 0.78    \\ \hline
			\textbf{$\overline{t}$(s)($N=5$, $M=100$)}  & 5.56      &1.79     & $ - $            & 1.87             \\\hline
            \textbf{ACR($N=5$, $M=500$)} & 0.89           & 0.84           &    $ - $     & 0.74   \\ \hline
			\textbf{$\overline{t}$(s)($N=5$, $M=500$)}  & 20.61      &2.18    & $ - $            & 2.26             \\\hline
   
		\end{tabular}%
	}
\end{table}

\section{Conclusion}
This paper investigates the movement optimization of multiple ABSs to maximize the ACR of mobile GUs in a site-specific environment. The problem is NP-hard in general and further complicated by the complex propagation environment and GU mobility. 
To tackle this challenging problem, we introduce the concept of GCM and formulate the problem into a BILP problem, for which a fast online algorithm is further proposed. 
Optimality bounds and complexity analysis are also provided. 
Numerical results demonstrate that our proposed algorithm achieves a high CR performance close to the upper bound obtained by SCIP, yet with significantly reduced running time. In addition, the algorithm also notably outperforms common benchmarks in the literature such as the K-means initiated evolutionary algorithm and one of the state-of-the-art DRL methods (TD3), in terms of CR performance and/or time efficiency.

\bibliography{IEEEabrv,bibliography}

\begin{thebibliography}{10}
\providecommand{\url}[1]{#1}
\csname url@samestyle\endcsname
\providecommand{\newblock}{\relax}
\providecommand{\bibinfo}[2]{#2}
\providecommand{\BIBentrySTDinterwordspacing}{\spaceskip=0pt\relax}
\providecommand{\BIBentryALTinterwordstretchfactor}{4}
\providecommand{\BIBentryALTinterwordspacing}{\spaceskip=\fontdimen2\font plus
\BIBentryALTinterwordstretchfactor\fontdimen3\font minus
  \fontdimen4\font\relax}
\providecommand{\BIBforeignlanguage}[2]{{%
\expandafter\ifx\csname l@#1\endcsname\relax
\typeout{** WARNING: IEEEtran.bst: No hyphenation pattern has been}%
\typeout{** loaded for the language `#1'. Using the pattern for}%
\typeout{** the default language instead.}%
\else
\language=\csname l@#1\endcsname
\fi
#2}}
\providecommand{\BIBdecl}{\relax}
\BIBdecl

\bibitem{LyuTWCHotspot}
J.~Lyu \emph{et~al.}, ``{UAV}-aided offloading for cellular hotspot,''
  \emph{IEEE Trans. Wireless Commun.}, vol.~17, no.~6, pp. 3988--4001, 2018.

\bibitem{2017-Lyu-Placement}
J.~Lyu, Y.~Zeng, R.~Zhang, and T.~J. Lim, ``Placement optimization of
  {UAV}-mounted mobile base stations,'' \emph{IEEE Commun. Lett.}, vol.~21,
  no.~3, pp. 604--607, Mar. 2017.

\bibitem{2016-Galkin-Deployment}
B.~Galkin, J.~Kibilda, and L.~A. DaSilva, ``Deployment of {UAV}-mounted access
  points according to spatial user locations in two-tier cellular networks,''
  in \emph{Wireless Days}, Mar. 2016, pp. 1--6.

\bibitem{2016-Mozaffari-Efficient}
M.~Mozaffari \emph{et~al.}, ``Efficient deployment of multiple unmanned aerial
  vehicles for optimal wireless coverage,'' \emph{IEEE Commun. Lett.}, vol.~20,
  no.~8, pp. 1647--1650, Aug. 2016.

\bibitem{8760267}
Z.~{Wang}, L.~{Duan}, and R.~{Zhang}, ``Adaptive deployment for {UAV}-aided
  communication networks,'' \emph{IEEE Trans. Wireless Commun.}, vol.~18,
  no.~9, pp. 4531--4543, Sep. 2019.

\bibitem{2014-Hourani-Optimal}
A.~Al-Hourani, S.~Kandeepan, and S.~Lardner, ``Optimal {LAP} altitude for
  maximum coverage,'' \emph{IEEE Wireless Commun. Lett.}, vol.~3, no.~6, pp.
  569--572, Dec. 2014.

\bibitem{2019-Lyu-RadioMap}
S.~Bi, J.~Lyu, Z.~Ding, and R.~Zhang, ``Engineering radio maps for wireless
  resource management,'' \emph{IEEE Wireless Commun.}, vol.~26, no.~2, pp.
  133--141, Apr. 2019.

\bibitem{10373821}
J.~Lyu, X.~Chen, J.~Zhang, and L.~Fu, ``Spatial deep learning for site-specific
  movement optimization of aerial base stations,'' \emph{IEEE Trans. Wireless
  Commun.}, pp. 1--1, Dec. 2023.

\bibitem{2020-Liu-Distributed}
C.~H. Liu, X.~Ma, X.~Gao, and J.~Tang, ``Distributed energy-efficient
  multi-{UAV} navigation for long-term communication coverage by deep
  reinforcement learning,'' \emph{IEEE Trans. Mobile Comput.}, vol.~19, no.~6,
  pp. 1274--1285, Jun. 2020.

\bibitem{2019-Liu-Reinforcement}
X.~Liu, Y.~Liu, and Y.~Chen, ``Reinforcement learning in multiple-{UAV}
  networks: Deployment and movement design,'' \emph{IEEE Trans. Veh. Technol.},
  vol.~68, no.~8, pp. 8036--8049, Aug. 2019.

\bibitem{2021-Zhang-Three-Dimension}
W.~Zhang, Q.~Wang, X.~Liu \emph{et~al.}, ``Three-dimension trajectory design
  for multi-{UAV} wireless network with deep reinforcement learning,''
  \emph{IEEE Trans. Veh. Technol.}, vol.~70, no.~1, pp. 600--612, Jan. 2020.

\bibitem{2020-Qiu-Placement}
J.~Qiu, J.~Lyu, and L.~Fu, ``Placement optimization of aerial base stations
  with deep reinforcement learning,'' in \emph{Proc. IEEE Int. Conf. Commun.
  (ICC)}, June 2020, pp. 1--6.

\bibitem{2020-Lyu-Codesign}
Z.~Lyu, C.~Ren, and L.~Qiu, ``Movement and communication co-design in
  multi-{UAV} enabled wireless systems via {DRL},'' in \emph{Proc. IEEE Int.
  Conf. Commun. China (ICCC)}, Dec. 2020, pp. 220--226.

\bibitem{CKMtutorial}
Y.~Zeng, J.~Chen, J.~Xu \emph{et~al.}, ``A tutorial on environment-aware
  communications via channel knowledge map for {6G},'' \emph{IEEE Commun.
  Surveys Tut.}, vol.~26, no.~3, pp. 1478--1519, Feb. 2024.

\bibitem{2019-Zhang-RadioMap-Path}
S.~Zhang and R.~Zhang, ``Radio map based path planning for cellular-connected
  {UAV},'' in \emph{Proc. IEEE Globecom}, Dec. 2019, pp. 1--6.

\bibitem{2021-Zeng-Navigation-RadioMap}
Y.~Zeng, X.~Xu, S.~Jin, and R.~Zhang, ``Simultaneous navigation and radio
  mapping for cellular-connected {UAV} with deep reinforcement learning,''
  \emph{IEEE Trans. Wireless Commun.}, vol.~20, no.~7, pp. 4205--4220, July
  2021.

\bibitem{10486853}
D.~Romero, P.~Q. Viet, and R.~Shrestha, ``Aerial base station placement via
  propagation radio maps,'' \emph{IEEE Trans. Commun.}, vol.~72, no.~9, pp.
  5349--5364, 2024.

\bibitem{2023-liu-UAV-Aided}
W.~Liu and J.~Chen, ``{UAV}-aided radio map construction exploiting environment
  semantics,'' \emph{IEEE Trans. Wireless Commun.}, vol.~22, no.~9, pp.
  6341--6355, Feb. 2023.

\bibitem{levie2021radiounet}
R.~Levie, {\c{C}}.~Yapar, G.~Kutyniok, and G.~Caire, ``{RadioUNet}: Fast radio
  map estimation with convolutional neural networks,'' \emph{IEEE Trans.
  Wireless Commun.}, vol.~20, no.~6, pp. 4001--4015, 2021.

\bibitem{zhen2022radio}
P.~Zhen, B.~Zhang, C.~Xie, and D.~Guo, ``A radio environment map updating
  mechanism based on an attention mechanism and siamese neural networks,''
  \emph{Sensors}, vol.~22, no.~18, p. 6797, 2022.

\bibitem{li2020simple}
X.~Li, C.~Sun, and Y.~Ye, ``Simple and fast algorithm for binary integer and
  online linear programming,'' \emph{Advances in Neural Information Processing
  Systems}, vol.~33, pp. 9412--9421, 2020.

\bibitem{BestuzhevaEtal2021ZR}
K.~Bestuzheva, M.~Besan{\c{c}}on, W.-K. Chen \emph{et~al.}, ``{The SCIP
  Optimization Suite 8.0},'' Zuse Institute Berlin, ZIB-Report 21-41, December
  2021.

\bibitem{3GPP}
3GPP-TR-36.777, ``{Enhanced LTE support for aerial vehicles},'' \emph{3GPP
  Technical Report}, Dec. 2017.

\bibitem{YangUAV}
H.~Xie, D.~Yang, L.~Xiao, and J.~Lyu, ``Connectivity-aware {3D UAV} path design
  with deep reinforcement learning,'' \emph{IEEE Trans. Veh. Technol.},
  vol.~70, no.~12, pp. 13\,022--13\,034, 2021.

\bibitem{luenberger1984linear}
Y.~Ye \emph{et~al.}, \emph{Linear and nonlinear programming}.\hskip 1em plus
  0.5em minus 0.4em\relax Springer, 1984.

\bibitem{gao2023solving}
W.~Gao, D.~Ge \emph{et~al.}, ``Solving linear programs with fast online
  learning algorithms,'' \emph{{in} Proc. Int. Conf. Machine Learning(ICML)},
  2023.

\end{thebibliography}
\end{document}